\documentclass[a4paper,12pt]{article}

\usepackage{hyperref}

\usepackage{amsmath}
\usepackage{graphicx}

\usepackage[margin=1.0in]{geometry}
\usepackage{authblk}

\newcommand{\dd}{\mathrm{d}}
\newcommand{\ee}{\mathrm{e}}
\newcommand{\ii}{\mathrm{i}}

\title{Nonminimally coupled scalar field in teleparallel gravity: boson stars}

\author{D.~Horvat}
\author{S.~Iliji\'c\footnote{Email: {\tt sasa.ilijic@fer.hr}}}
\author{A.~Kirin}
\author{Z.~Naran\v ci\'c}

\affil{University of Zagreb, Faculty of Electrical Engineering and Computing,\\
   Department of Applied Physics, Unska 3, HR-10\,000 Zagreb, Croatia}

\date{\today}

\begin{document}

\maketitle 

%%%%%%%%%%%%%%%%%%%%%%%%%%%%%%%%%%%%%%%%%%%%%%%%%%%%%%%%%%%%%%%%%%%%%%%%%%%%%%%

\begin{abstract}
We study the nonminimally coupled complex scalar field
within the framework of teleparallel gravity.
Coupling of the field nonminimally to the torsion scalar
destroys the Lorentz invariance of the theory
in the sense that the resulting equations of motion
depend on the choice of a tetrad.
For the assumed static spherically symmetric spacetime,
we find a tetrad which leads to a self-consistent set of equations,
and we construct the self-gravitating configurations
of the scalar field---boson stars.
The resulting configurations develop anisotropic principal pressures
and satisfy the dominant energy condition.
An interesting property of the configurations
obtained with sufficiently large field-to-torsion coupling constant
is the outwardly increasing energy density,
followed by an abrupt drop towards the usual asymptotic tail.
This feature is not present in the boson stars with the
field minimally or nonminimally coupled to the curvature scalar,
and therefore appears to be a torsion--only effect.
\end{abstract}

%%%%%%%%%%%%%%%%%%%%%%%%%%%%%%%%%%%%%%%%%%%%%%%%%%%%%%%%%%%%%%%%%%%%%%%%%%%%%%%

\section{Introduction \label{sec:intro}}

%%%%%%%%%%%%%%%%%%%%%%%%%%%%%%%%%%%%%%%%%%%%%%%%%%%%%%%%%%%%%%%%%%%%%%%%%%%%%%%

Teleparallel gravity \cite{aldrovandi,maluf}
is a gravity theory based on spacetime torsion,
instead of curvature on which
standard general relativity (GR) is based.
The dynamical quantities of teleparallel gravity are tetrad fields 
that determine the orthonormal basis of the tangent space
at every spacetime point.
In terms of the tetrad fields
one constructs the curvature-less Weitzenb\"ock connection,
which is used instead of the torsion-less Levi--Civita connection of GR.
Writing the gravitational action as
   \begin{equation}
   S = \int \frac{T}{2k} \, h \, \dd^4 x,
   \end{equation}
where $T$ is the suitably defined torsion scalar,
and $h\,\dd^4 x$ is the proper volume element,
the equations of motion
equivalent to those of GR are obtained,
and this particular variant of the theory is known
as the teleparallel equivalent of general relativity (TEGR).

The gravitational sector of TEGR is Lorentz invariant
in the sense that any choice of the tetrad fields
leads to the same equations of motion.
Extensions of this theory, such as the $f(T)$-gravity,
or direct coupling of matter fields to the torsion scalar,
disrupt the Lorentz invariance of the equations of motion.
In principle, the equations of motion must be employed
to fix the extra degrees of freedom (boost and rotation)
contained in the choice of the tetrad fields \cite{barrow1,barrow2}.
Regardless of these difficulties,
there is growing interest in the $f(T)$ theory
and/or teleparallel gravity with nonminimal coupling.
Most applications are in the fields of cosmology
\cite{barrow3,sadjadi,harkolobo,mubasher}
and dark energy models
\cite{gengleesaridakis,geng1,gujaleegeng,weihao,kucudark},
while in the spherical symmetry
one finds \cite{boehmerbh,boeh1,ulhoa1,Aftergood:2014wla}.
Nonminimal coupling of the energy-momentum tensor to torsion
has recently been shown to offer a possibility
of detecting torsion experimentally \cite{obuk2,obuk1}.

In this paper we consider the massive complex scalar field
nonminimally coupled to the torsion scalar.
We investigate the possibility of forming static
spherically symmetric self-gravitating configurations,
analogous to the boson stars that have been thoroughly investigated
within the standard curvature theory (GR).
Boson stars
first appeared in \cite{KaupKleinGordonGeon} and \cite{RuffiniBonazzola},
and despite of the fact that their existence in nature is unconfirmed,
they have had their roles in various contexts of physics.
At the astrophysical front, boson stars were considered
as models of massive galactic centres \cite{torres2000},
as black-hole mimickers \cite{guzman},
as galactic dark matter halos explaining the rotation curves \cite{Lee:1995af},
etc.
They also turned out to be useful tools in mathematical relativity
where critical phenomena and gravitational collapse
could be studied \cite{hawchop}.
The reviews written over the past decades
contain the overview of the field
\cite{liddlemadsen,jetzer,schunck,liebling}.
In particular, boson stars with nonminimal coupling
of the scalar field to the curvature scalar
have first been considered in \cite{vanderbij},
and some of their properties were further investigated
in \cite{photonspheres,horvatmarunovic}.

The paper is organized as follows:
In section~\ref{sec:notation} we briefly go
through the basic notions of the teleparallel gravity,
mainly to establish the notation and to introduce
the needed quantities.
In section~\ref{sec:nmct} we derive the general equations of motion
for the scalar field nonminimally coupled to the torsion scalar.
In section~\ref{sec:stars} we restrict the analysis to spherical symmetry.
We find an appropriate tetrad, derive the equations of motion,
construct static spherically symmetric solutions,
and discuss their properties.
We sum up in section~\ref{sec:concl}.
Geometrized units, $G_{\mathrm{N}} = 1 = c$, are used throughout the paper.

%%%%%%%%%%%%%%%%%%%%%%%%%%%%%%%%%%%%%%%%%%%%%%%%%%%%%%%%%%%%%%%%%%%%%%%%%%%%%%%

\section{Teleparallel gravity: notation and conventions\label{sec:notation}}

%%%%%%%%%%%%%%%%%%%%%%%%%%%%%%%%%%%%%%%%%%%%%%%%%%%%%%%%%%%%%%%%%%%%%%%%%%%%%%%

Teleparallel gravity can be formulated
in terms of the tetrad fields, $h_a{}^{\mu}$,
which determine the local Lorentz frame at every spacetime point.
Latin indices run over the Lorentz frame coordinates,
and Greek indices run over the spacetime coordinates.
Tetrad fields obey the following well-known relations,
   \begin{equation}
   \eta_{ab} = h_a{}^{\mu} h_b{}^{\nu} g_{\mu\nu}, \qquad
   g_{\mu\nu} = h^a{}_{\mu} h^b{}_{\nu} \eta_{ab}, \qquad
   h^a{}_{\mu} h_b{}^{\mu} = \delta^a_b, \qquad
   h^a{}_{\alpha} h_a{}^{\beta} = \delta^{\beta}_{\alpha},
   \end{equation}
where $\eta_{ab}=\mathrm{diag}(-,+,+,+)$
is the metric in the Lorentz frame,
and $g_{\mu\nu}$ is the spacetime metric tensor.
It is important to emphasize
that while the tetrad fields fully determine the spacetime metric,
the converse is not true;
at every spacetime point, there is a six-fold infinity of tetrads,
mutually related by the spacetime-dependent Lorentz transformations
(these involve three boost and three rotation parameters),
all giving raise to the same spacetime metric.
Instead of the torsion-less Levi--Civita connection of GR,
here denoted with $\Gamma^{\alpha}{}_{\beta\gamma}$,
one adopts the curvature-less Weitzenb\"ock connection,
here denoted with the tilded symbol
   \begin{equation} \label{eq:weitz}
   \tilde \Gamma^{\alpha}{}_{\beta\gamma}
      \equiv h_a{}^{\alpha} h^a{}_{\beta,\gamma},
   \end{equation}
and proceeds to define the torsion tensor and the torsion vector
   \begin{equation} \label{eq:ttensor}
   \tilde T^{\alpha}{}_{\beta\gamma} \equiv
       - 2 \tilde \Gamma^{\alpha}{}_{[\beta\gamma]}
   = \tilde \Gamma^{\alpha}{}_{\gamma\beta}
        - \tilde \Gamma^{\alpha}{}_{\beta\gamma}, \qquad
   \tilde T_{\alpha} \equiv \tilde T^{\mu}{}_{\alpha\mu} .
   \end{equation}
In the above expressions, and in what follows,
the quantities derived using the Weitezenb\"ock connection,
and belonging to the formalism of the teleparallel gravity,
are denoted with the tilde,
while those derived with the Levi--Civita connection
of the standard GR are not tilded.
The contortion tensor is defined as the difference
between the Weitzenb\"ock and the Levi--Civita connections,
and can be written in terms of the torsion tensor
   \begin{equation} \label{eq:ktensor}
   \tilde K_{\alpha\beta\gamma} \equiv
   \tilde \Gamma^{\alpha}{}_{\beta\gamma} - \Gamma^{\alpha}{}_{\beta\gamma}
   = \frac12 \left( \tilde T_{\alpha\gamma\beta}
     + \tilde T_{\beta\alpha\gamma} + \tilde T_{\gamma\alpha\beta} \right) ,
   \end{equation}
and the so-called modified torsion tensor is defined as
   \begin{equation} \label{eq:stensor}
   \tilde S_{\alpha\beta\gamma} \equiv \tilde K_{\beta\gamma\alpha}
      + g_{\alpha\beta} \, \tilde T_{\gamma}
      - g_{\alpha\gamma} \, \tilde T_{\beta}.
   \end{equation}
(The above definitions imply the following properties:
$ \tilde T_{\alpha(\beta\gamma)} = 0 $,
$ \tilde K^{\mu}{}_{\alpha\mu} = - \tilde T_{\alpha} $,
$ \tilde K_{(\alpha\beta)\gamma} = 0 $,
$ \tilde S_{\alpha(\beta\gamma)} = 0 $.)
Finally, the torsion scalar is defined as
   \begin{equation} \label{eq:tscalar}
   \tilde T \equiv
   \frac12 \tilde S_{\alpha\beta\gamma} \tilde T^{\alpha\beta\gamma}
   = \frac14 \tilde T_{\alpha\beta\gamma} \tilde T^{\alpha\beta\gamma}
     + \frac12 \tilde T_{\alpha\beta\gamma} \tilde T^{\gamma\beta\alpha}
     - \tilde T_{\alpha} \tilde T^{\alpha}.
   \end{equation}
The torsion scalar is a generally covariant scalar,
which means that it is invariant under infinitesimal
spacetime coordinate transformations,
$x^{\alpha} \to x^{\alpha}+\epsilon^{\alpha}(x)$,
but it is \emph{not} a local Lorentz scalar,
since it is not invariant with respect to spacetime-dependent
(local) Lorentz transformations of the tetrad, or in other words,
it depends on the particular choice of the tetrad \cite{barrow1}.

It can be shown that the torsion scalar, $\tilde T$,
and the Ricci curvature of the spacetime, $R$, are related by
   \begin{equation} \label{eq:tr}
   R = - \tilde T - \frac{2}{h} \partial_{\mu} (h \tilde T^\mu) ,
   \end{equation}
which means that they differ, apart from the sign,
only in the total divergence of a vector field.
Since the total divergence does not affect the variation of the action,
it follows that one can replace the Ricci curvature scalar
in the Einstein--Hilbert action of GR with $-\tilde T$,
i.e.\ write the action as
   \begin{equation} \label{eq:tegraction}
   S = \int \mathrm{d} x^4 \, h
       \left( - \frac{\tilde T}{2k} + L_{\mathrm{matter}} \right) ,
   \end{equation}
where $h=\det(h_a{}^{\alpha}) = \sqrt{-\det{g_{\alpha\beta}}}$,
$k=8\pi$ is the coupling constant, and $L_{\mathrm{matter}}$
is the Lagrangian involving the matter fields,
and obtain the equations of motion that are equivalent to those of GR.
Therefore, although the action (\ref{eq:tegraction})
is not Lorentz invariant
(since $\tilde T$ is itself not Lorentz invariant),
the equations of motion are Lorentz invariant.
The resulting theory is known
as the TEGR.
The Lorentz invariance of TEGR is lost already in its simplest extensions
such as the one we are considering in the next section.

%%%%%%%%%%%%%%%%%%%%%%%%%%%%%%%%%%%%%%%%%%%%%%%%%%%%%%%%%%%%%%%%%%%%%%%%%%%%%%%

\section{Torsion coupled scalar field \label{sec:nmct}}

%%%%%%%%%%%%%%%%%%%%%%%%%%%%%%%%%%%%%%%%%%%%%%%%%%%%%%%%%%%%%%%%%%%%%%%%%%%%%%%

The action involving the complex scalar field $\phi$ coupled to torsion,
that most closely resembles the well-known case of
nonminimal coupling of the scalar field to the curvature scalar,
can be written as $S = \int \mathcal L \; \mathrm{d}^4 x $, where
   \begin{equation} \label{eq:lag}
   \mathcal L =
       \frac{h}{2k} \big(1 + 2k\xi \phi^*\phi\big) (- \tilde T)
       - h \left( \frac12 g^{\alpha\beta}
       \big( \phi_{,\alpha}^* \phi_{,\beta}
       + \phi_{,\beta}^* \phi_{,\alpha} \big)
       + \mu^2 \phi^*\phi \right)
   \end{equation}
is the Lagrangian density and $\xi$ is the field-to-torsion coupling constant.
The scalar field is taken to be massive, $\mu$ being the mass parameter,
while for simplicity we are not introducing the field self-interaction.

Variation of the action with respect to the tetrad
leads to the Euler--Lagrange equation
$ \partial_{\mu} ( { \partial \mathcal{L} }
   / { \partial( \partial_\mu h^a{}_\nu ) })
= { \partial \mathcal{L} } / { \partial( h^a{}_\nu ) } $.
Using
   \begin{equation}
   \frac{\partial h}{\partial(h^a{}_\nu)} = h h_a{}^\nu, \qquad
   \frac{\partial \tilde T}{\partial(h^a{}_\nu)}
      = - 2 h_a{}^\gamma \tilde T_{\alpha\beta\gamma}
                         \tilde S^{\alpha\beta\nu}, \qquad
   \frac{\partial \tilde T}{\partial(\partial_\mu h^a{}_\nu)}
      = - 2 \tilde S_a{}^{\nu\mu}
   \end{equation}
(for a detailed derivation of the above relations,
see e.g.\ appendix C of~\cite{aldrovandi}),
one obtains
   \begin{alignat}{1}
   & \partial_{\mu} \Big( \frac{h}{2k} \big(1 + 2k\xi \phi^*\phi\big)
     (2 \tilde S_{a}{}^{\nu\mu}) \Big) =
   \notag \\
   & \quad h_a{}^{\nu} \mathcal L
   + \frac{h}{2k} \big(1 + 2k\xi \phi^*\phi\big)
         (2 \tilde T_{\alpha\beta a} \tilde S^{\alpha\beta\nu})
   + \frac{h}{2}
       \big( g^{\nu\beta} h_a{}^{\alpha} + g^{\nu\alpha} h_a{}^{\beta} \big)
       \big( \phi_{,\alpha}^* \phi_{,\beta}
             + \phi_{,\beta}^* \phi_{,\alpha} \big) .
   \end{alignat}
Contracting with $h^a{}_{\rho}$ and multiplying by $k/h$,
the above equation of motion
can be written in the form of the Einstein equation,
   \begin{equation} \label{eq:albert}
   G^{\nu}{}_{\rho} = k T^{\nu}{}_{\rho},
   \end{equation}
where
   \begin{equation} \label{eq:einten}
   G^{\nu}{}_\rho = \frac12 \tilde T \delta_{\rho}^{\nu}
      - \tilde T_{\alpha\beta\rho} \tilde S^{\alpha\beta\nu}
          + \frac1h h^a{}_{\rho} \partial_{\mu}
                \big( h \tilde S_{a}{}^{\nu\mu} \big)
   % = R_{\rho}^{\nu} - \frac12 \delta_{\rho}^{\nu} R
   \end{equation}
coincides with the Einstein tensor of GR
($G^{\nu}{}_{\rho} = R_{\rho}^{\nu} - \frac12 \delta_{\rho}^{\nu} R$,
obtained using the Levi--Civita connection),
and the energy--momentum tensor is given by
   \begin{equation} \label{eq:Tgen}
   T_{\mu\nu} = \frac{
      \phi_{,\mu}^* \phi_{,\nu} + \phi_{,\nu}^* \phi_{,\mu}
      - g_{\mu\nu} \Big(
         \frac12 g^{\alpha\beta} \big( \phi_{,\alpha}^* \phi_{,\beta}
         + \phi_{,\beta}^* \phi_{,\alpha} \big) + {\mu}^2 \phi^*\phi \Big)
      - {2 \xi} \tilde S_{\nu\mu}{}^{\alpha} \partial_{\alpha} ( \phi^*\phi )
   }{ 1 + 2k\xi \phi^* \phi } .
   \end{equation}
For $\xi=0$, the above expression for the energy--momentum tensor
reduces to what one expects for the minimally coupled field in GR,
while with $\xi\ne0$,
comparing it with the energy--momentum tensor
for the scalar field nonminimally coupled to the curvature scalar,
reveals the difference only in the last term in the numerator.
In the case of the curvature--coupled field, this term reads
   \begin{equation}
   - {2 \xi} g_{\mu\nu} \nabla^{\alpha} \nabla_{\alpha} \phi^* \phi
   + {2 \xi} \nabla_{\mu} \nabla_{\nu} \phi^* \phi ,
   \end{equation}
which involves second order derivatives of the field,
while the expression (\ref{eq:Tgen})
involves only the first order derivatives.

Variation of the action with respect to the field
gives the equation of motion for the scalar field
   \begin{equation} \label{eq:kleingordon}
   \nabla_{\mu} \nabla^{\mu} \phi
   = ( \xi \tilde T + \mu^2 ) \phi,
   \end{equation}
which for $\xi=0$ reduces to the Klein--Gordon equation.
Let us also note that, due to the invariance of the action
with respect to global transformation of the field,
$\phi \to \ee^{\ii \epsilon} \phi$,
we have the conserved current,
$j_{\mu} = \ii ( (\nabla_\alpha \phi^*)\phi - (\nabla_\alpha \phi)\phi^* )$,
and therefore the conserved charge
that can be interpreted as the particle number, $N$.

%%%%%%%%%%%%%%%%%%%%%%%%%%%%%%%%%%%%%%%%%%%%%%%%%%%%%%%%%%%%%%%%%%%%%%%%%%%%%%%

\section{Boson stars \label{sec:stars}}

%%%%%%%%%%%%%%%%%%%%%%%%%%%%%%%%%%%%%%%%%%%%%%%%%%%%%%%%%%%%%%%%%%%%%%%%%%%%%%

As the specific example of the behaviour
of the torsion--coupled scalar field
we will consider the possibility of forming
static spherically symmetric self-gravitating structures,
which we will call boson stars.

The line element of the static spherically symmetric spacetime
can be written using the coordinates $x^{\mu}=(t,r,\vartheta,\varphi)$ as
   \begin{equation} \label{eq:ds2}
   \dd s^2 = - \ee^{2\Phi(r)} \, \dd t^2
             + \ee^{2\Lambda(r)} \, \dd r^2
             + r^2 \, \dd\Omega^2  ,
   \end{equation}
where $\Phi$ and $\Lambda$ are the two $r$-dependent metric profile functions,
and $\dd\Omega^2 = \dd\vartheta^2 + \sin^2\theta \, \dd\varphi^2$
is the metric on the unit sphere.
For the above metric,
the components of the Einstein tensor (\ref{eq:einten})
can be obtained from the standard curvature tensors,
i.e.\ without making any reference to the tetrad.
It's non-zero components are
   \begin{alignat}{1}
   G^t{}_t & = r^{-2} \big( \mathrm{e}^{-2\Lambda}(1-2r\Lambda')-1\big),
      \label{eq:einten00} \\
   G^r{}_r & = r^{-2} \big( \mathrm{e}^{-2\Lambda}(1+2r\Phi')-1\big), \\
   G^\vartheta{}_\vartheta = G^\varphi{}_\varphi & = 
      r^{-2} \mathrm{e}^{-2\Lambda} \big(
         (r\Phi' - r\Lambda')(1 + r\Phi') + r^2 \Phi'' \big),
      \label{eq:einten22}
   \end{alignat}
where explicit notation of the $r$-dependencies is omitted,
and prime ($'$) denotes the $r$-derivatives.

To compute the energy-momentum tensor (\ref{eq:Tgen}) in the present context,
we first adopt the usual time--stationary ansatz
for the complex scalar field in spherical symmetry
   \begin{equation}
   \phi(t,r) = \frac1{\sqrt k} \, \sigma(r) \, \ee^{ - \ii \omega t},
   \end{equation}
where $\sigma(r)$ is the real field profile function
and the constant $\omega$ is the frequency.
The terms in the numerator of (\ref{eq:Tgen}),
apart from the last one,
can be calculated without the reference to the tetrad,
and they yield the well-known
contributions to the diagonal of the energy--momentum tensor.
The last term in the numerator of (\ref{eq:Tgen})
involves the modified torsion tensor (\ref{eq:stensor}), and therefore,
in order to complete the calculation of the energy--momentum tensor,
the tetrad must be chosen.

As our first choice for the tetrad,
we make use of the `square root of the metric tensor recipe',
which gives the `diagonal' tetrad,
   \begin{equation} \label{eq:tetrad1}
   h^a{}_{\mu} = \mathrm{diag}(\ee^{\Phi},\ee^{\Lambda},r,r\sin\vartheta) .
   \end{equation}
For the above tetrad we compute the Weitzenb\"ock connection (\ref{eq:weitz}),
the suite of torsion-related tensors (\ref{eq:ttensor})--(\ref{eq:tscalar}),
and use the modified torsion tensor (\ref{eq:stensor})
to obtain the energy--momentum tensor (\ref{eq:Tgen}).
It is immediately revealed that
the last term in the numerator of (\ref{eq:Tgen})
gives non-diagonal and non-symmetrical
contribution to the energy--momentum tensor.
In particular, we obtain
   \begin{equation}
   T_{r\vartheta}
   = \xi \frac{4}{kr^2} \ee^{2\Lambda} \sigma \sigma' \cot\vartheta,
   \qquad
   T_{\vartheta r} = 0,
   \end{equation} 
which is inconsistent with the structure of the Einstein equation,
except if $\xi=0$, which sets one back to the minimal coupling case.
At this point we conclude that the `diagonal' tetrad (\ref{eq:tetrad1})
is not suitable for the present problem.

As another choice of the tetrad we take
   \begin{equation} \label{eq:tetrad2}
   h^a{}_{\mu} = \left( \begin{array}{cccc}
   \ee^{\Phi} & 0 & 0 & 0 \\
   0 & \ee^{\Lambda} \sin\vartheta \cos\varphi
     &  r \sin\vartheta \cos\varphi & -r \sin\vartheta \sin\varphi \\
   0 & \ee^{\Lambda} \sin\vartheta \sin\varphi
     &  r \sin\vartheta \cos\varphi &  r \sin\vartheta \cos\varphi \\
   0 & \ee^{\Lambda} \cos\vartheta & -r \sin\vartheta             &  0
   \end{array}
   \right),
   \end{equation}
which is related to (\ref{eq:tetrad1}) by a spacetime-dependent rotation.
Repeating the procedure,
we compute the new suite of torsion-related tensors\footnote{%
For example, with the `diagonal' tetrad (\ref{eq:tetrad1})
for the torsion scalar (\ref{eq:tscalar}) we obtain
$\tilde T=-2r^{-2}\ee^{-2\Lambda}(1+2r\Phi')$,
while with the `rotated' tetrad (\ref{eq:tetrad2}) we obtain
$\tilde T=-2r^{-2}\ee^{-2\Lambda}(\ee^\Lambda-1) (\ee^{\Lambda}-1-2r\Phi')$.
For comparison, the Ricci curvature scalar
corresponding to the metric (\ref{eq:ds2}) is
$R=2r^{-2} \ee^{-2\Lambda} ( \ee^{2\Lambda} - 1
+ (r\Lambda' - r\Phi')(2+r\Phi') - r^2\Phi'' )$.}
and we find that in this case the energy--momentum tensor is diagonal,
as is the Einstein tensor.
Writing $T^{\mu}{}_{\nu} = \mathrm{diag}(-\rho,p,q,q)$,
allows one to identify the non-zero components
of the energy--momentum tensor as the energy density
   \begin{equation} \label{eq:rho}
   \rho = - T^t{}_t = \frac{
      (\ee^{-2\Phi}\omega^2 + \mu^2) \sigma^2
    + \ee^{-2\Lambda} \sigma'^2
    + 8\xi r^{-1} \ee^{-2\Lambda} (\ee^{\Lambda}-1) \sigma \sigma'
   }{k(1+2\xi\sigma^2)},
   \end{equation}
the radial pressure
   \begin{equation} \label{eq:ppp}
   p = T^r{}_r = \frac{
      (\ee^{-2\Phi}\omega^2 - \mu^2) \sigma^2
    + \ee^{-2\Lambda} \sigma'^2
   }{k(1+2\xi\sigma^2)},
   \end{equation}
and the transverse pressure
   \begin{equation} \label{eq:qqq}
   q = T^\vartheta{}_\vartheta = T^\varphi{}_\varphi = \frac{
      (\ee^{-2\Phi}\omega^2 - \mu^2) \sigma^2
    - \ee^{-2\Lambda} \sigma'^2
    + 4 \xi r^{-1} \ee^{-2\Lambda} (\ee^{\Lambda}-1 - r \Phi') \sigma \sigma'
   }{k(1+2\xi\sigma^2)}.
   \end{equation}
With the components of the Einstein tensor,
given by (\ref{eq:einten00})--(\ref{eq:einten22}),
and the components of the energy momentum tensor
obtained with the `rotated tetrad' (\ref{eq:tetrad2}),
given by (\ref{eq:rho})--(\ref{eq:qqq}),
we find that the Einstein equation (\ref{eq:albert})
consists of three independent ordinary differential equations,
involving three unknown functions, $\Phi$, $\Lambda$ and $\sigma$,
and one unknown constant $\omega$.
As the additional test of the internal consistency of our equations,
we verified that the field equation of motion (\ref{eq:kleingordon}) is,
in this context,
equivalent to the conservation condition $\nabla^\mu T_{\mu\nu}=0$.
We therefore adopt the equations of motion
obtained using the `rotated' tetrad (\ref{eq:tetrad2})
as our choice for the analysis of the boson stars with torsion--coupled field.

In order to construct the solutions to the Einstein equation discussed above,
we rely on the numerical procedures.
We do so by posing the boundary value problem (BVP),
where as the boundaries we take the centre of the symmetry, $r=0$,
and spatial infinity, $r=\infty$.
This choice of the outer boundary allows the field to take up all space,
and is common in the context of boson stars.
However, it is not completely general,
as one can also require that the field, at some finite $r=R$,
behaves in such a way that the interior solution
can be smoothly joined with the exterior vacuum solution.
Such interior solutions
have been recently obtained within the standard theory \cite{compactbstar},
and are called compact boson stars.
They have the sharply defined surface radius $R$,
while in the solutions we are to construct,
the vacuum state is reached only asymptotically as $r\to\infty$.
The boundary conditions reflect the expected behaviour
of the metric profile functions, $\Phi$ and $\Lambda$,
and the field profile function, $\sigma$, at the boundaries.
The Einstein equations involve $\Phi''$, $\Lambda'$ and $\sigma'$
as the highest order derivatives of the unknown functions,
but it turns out to be convenient to differentiate the $({}_r^r)$-component
of the Einstein equation and eliminate $\Phi''$ from the system,
ending up with $\Phi'$, $\Lambda'$ and $\sigma''$
as the highest order derivatives.
(This step is equivalent to using the field-equation (\ref{eq:kleingordon}),
or the conservation condition $\nabla^\alpha T_{\alpha\mu}=0$.)
As the system also involves the unknown constant $\omega$,
which has the role of the eigenvalue,
we extend it by adding the differential equation $\omega'=0$.
The summed order of the differential equations in the system equals five,
requiring five boundary conditions, which we chose as
   \begin{equation} \label{eq:bc}
   \Phi(\infty) = 0, \qquad
   \Lambda(0) = 0, \qquad
   \sigma(0) = \sigma_0, \qquad
   \sigma'(0) = 0, \qquad
   \sigma(\infty) = 0.
   \end{equation}
The boundary condition $\Phi(\infty) = 0$
affects only the global scaling of the time coordinate,
and this particular value is chosen
in accord with the usual form of the flat metric at spatial infinity.
The condition $\Lambda(0)=0$ follows from the requirement
that the energy density is finite at $r=0$.%
\footnote{
One can see this by writing the metric component as
$ g_{rr} = \ee^{2\Lambda} = (1-2m/r)^{-1}$,
where $m(r)$ is the usual `mass function'.
As the Einstein equations imply $m' = 4\pi r^2 \rho $,
requiring that $\rho$ is finite as $r\to0$ leads to
the boundary condition $\Lambda(0)=0$.
As another option, requiring that $m(r)$ becomes constant as $r\to\infty$
leads to $\Lambda(\infty)=0$, which could also be used as a boundary condition.
}
The condition $\sigma(0)=\sigma_0$
introduces the central value of the field profile function
which is used as the parameter to generate families of solutions
corresponding to fixed values of $\xi$ and $\mu$.
The condition $\sigma'(0)=0$ ensures
that the second derivative of $\sigma$ at $r=0$ remains finite
(diverging $\sigma''$ would,
by virtue of the field equation (\ref{eq:kleingordon}),
imply the divergence of the torsion scalar $\tilde T$).
The condition $\sigma(\infty)=0$
ensures the vanishing of the energy density at spatial infinity.
As an additional restriction,
we are only considering the configurations
in which the field profile function has no nodes.
Using the radial variable $x=r/(r+1)$,
the problem is formulated on the compact domain,
and the solutions are constructed with the
collocation--algorithm based code \textsc{colsys}
\cite{COLSYS}.
The primary advantage of the BVP approach is
that it constructs the solution over the whole domain,
automatically providing the eigenvalue $\omega$.
Of course, each solution obtained through the BVP procedure
can be double-checked by carrying out
the initial value problem (IVP) integration, starting at $r=0$,
taking $\Lambda(0)=0$, $\Phi(0)=\Phi_0$,
where $\Phi_0$ is obtained through BVP,
$\sigma(0)=\sigma_0$, $\sigma'(0)=0$,
and $\omega$ as obtained through BVP, as the initial values.
The IVP integration reveals high sensitivity on the value of $\omega$
(or equivalently $\Phi_0$).
Slight departures of $\omega$ from the correct value (obtained through BVP)
make the field function $\sigma$ diverge well before $x=1$ is reached.
Although the IVP approach can, in principle,
be used to confine the correct value of $\omega$
(see e.g.\ the appendix of~\cite{RuffiniBonazzola}), 
we have found the BVP approach simpler to use
and also numerically more stable.

Apart from the metric and the field profile functions,
the quantities of interest are the total mass, $M$,
and the total particle number, $N$, of the boson star,
as the binding energy of a star can be defined as the
difference between its total mass
and the rest energy of the particles dispersed at infinity.
In terms of $M$, $N$, and the field mass parameter $\mu$,
the binding energy is given by
   \begin{equation} \label{eq:ebind}
   E_b = M - \mu N.
   \end{equation}
In order to evaluate the total mass of a star
in a spherically symmetric spacetime
it is convenient to write $g_{rr}$ in terms of the `mass function' $m(r)$,
   \begin{equation}
   g_{rr} = \ee^{2\Lambda(r)} = (1-{2m(r)}/{r})^{-1},
   \end{equation}
since the asymptotic value of $m(r)$
as $r \to \infty$ is the total mass of the star, $M$.
(The ratio $2m(r)/r<1$ is known as the compactness function
and is a measure of the compactness of the object at certain $r$.)
The total particle number, $N$,
is the integral of the time-component
of the conserved current over the spatial slice
   \begin{equation}
   N = \int \dd^3 x \sqrt{-g} j^0
     = \int_0^{\infty} 8 \pi r^2
       \ee^{\Lambda-\Phi} \omega k^{-1 }\sigma^2 \, \dd r,
   \end{equation}
and can be evaluated after the solution has been obtained.

\begin{figure}
\begin{center}
\includegraphics{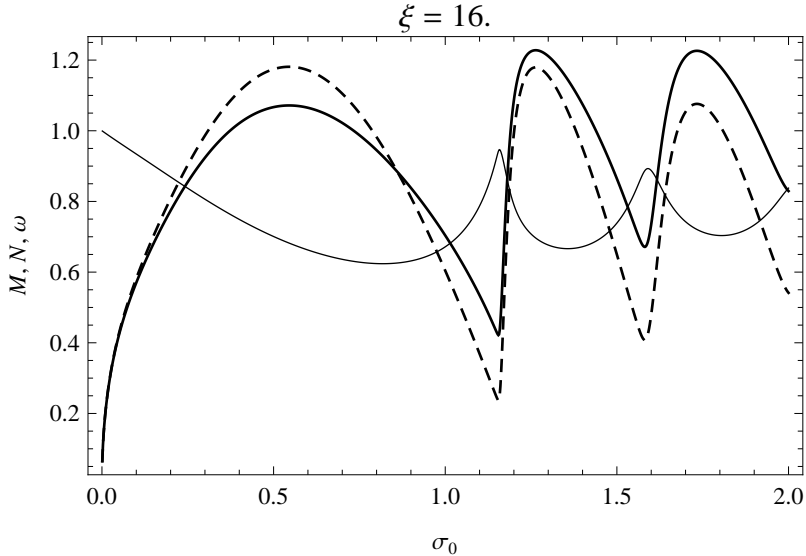}
\end{center}
\caption{\label{fig:one}
Boson stars with field--to--torsion coupling $\xi=16$:
total mass $M$
(thick line, in units of $M_{\mathrm{Pl}}^2/\mu$),
particle number $N$
(dashed line, in units of $M_{\mathrm{Pl}}^2/\mu^2$),
and the frequency $\omega$
(thin line, in units of $\mu/M_{\mathrm{Pl}}^2$),
are shown for a range of central values
of the field profile function $\sigma(0)$.}
\end{figure}

\begin{figure}
\begin{center}
\includegraphics{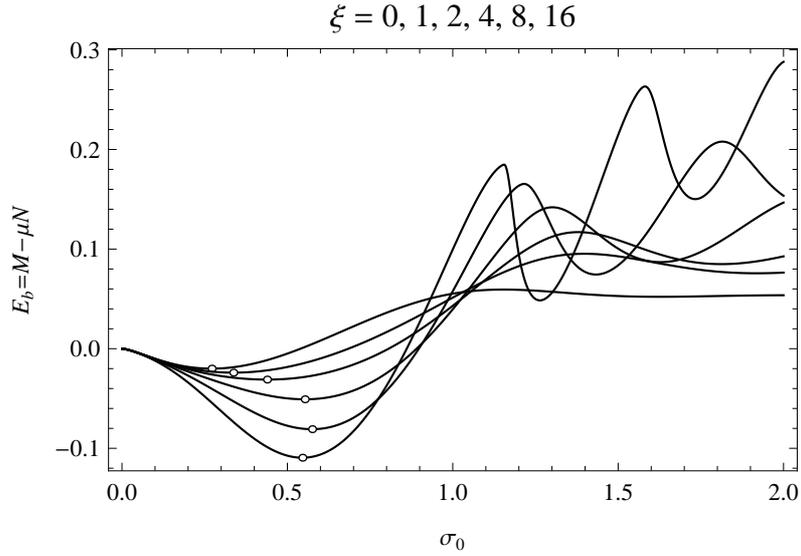}
\end{center}
\caption{\label{fig:two}
Binding energy of boson stars in units of $M_{\mathrm{Pl}}^2/\mu$
with field--to--torsion coupling $\xi=0,1,2,4,8,16$,
and a range of values of $\sigma_0$.
Critical solutions are indicated with circles.
The minima become deeper as $\xi$ increases.}
\end{figure}

The central value of the field profile function, $\sigma_0$,
can be used to parametrize the spectrum of solutions
corresponding to the chosen value of the field mass, $\mu$,
and the field-to-torsion coupling constant, $\xi$.
In figure~\ref{fig:one} we show the behaviour of the total mass, $M$,
the particle number, $N$,
and the value of the frequency $\omega$,
as the $\sigma_0$ increases,
in solutions obtained with $\mu=1$ and $\xi=16$.
One notices the clearly pronounced coinciding maxima in $M$ and $N$,
as well as the very sharp minima in between of them.
Similar oscillations in $M$ and $N$ are found with $\xi=0$ (minimal coupling)
where the solution corresponding to the first maximum in the mass
(as $\sigma_0$ increases) is referred to as the critical solution
because it coincides
with the onset of the dynamical instability of the star
in the usual curvature theory \cite{gleiserwatkins}.
At present, it is not possible to tell
whether the solutions with the torsion-coupled field
suffer from the same property at the first maximum of the total mass,
but we will nonetheless refer to them as the critical solutions.
Figure~\ref{fig:two} shows
the binding energy (\ref{eq:ebind}) for several values of $\xi$
over a range of values of $\sigma_0$.
We see that all critical solutions have negative binding energy,
which is a property of gravitationally bound systems,
but has no direct implications on the stability of the stars.

\begin{figure}
\begin{center}
\includegraphics{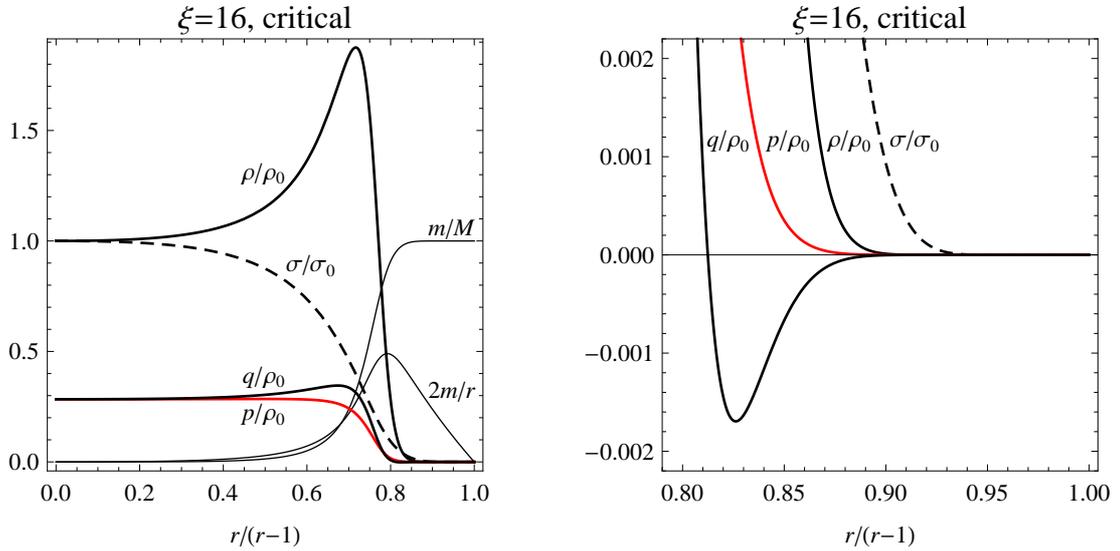}
\end{center}
\caption{\label{fig:threefour}
Critical solution with $\xi=16$:
field profile function $\sigma$
relative to its value at the centre of the star (dashed line),
the energy density $\rho$,
the radial pressure $p$ and the transverse pressure $q$
relative to the central value of the energy density (thick solid lines),
the mass function $m$ relative to the total mass of the star
and the compactness function $2m/r$ (thin solid lines).
The panel to the right shows the close-up of the
the asymptotic behaviour of $\sigma$, $\rho$, $p$ and $q$.}
\end{figure}

In figure~\ref{fig:threefour},
we show the scalar field profile function $\sigma$,
the components of the energy--momentum tensor,
the mass function $m$, and the compactness function $2m/r$,
for the critical solution with $\xi=16$.
While the field profile is outwardly decreasing,
as in the familiar case of the boson star with the field
minimally \cite{liddlemadsen}
or nonminimally \cite{vanderbij,horvatmarunovic}
coupled to the curvature scalar,
a unexpected and interesting feature that we find here
is the outwardly increasing energy density,
followed by the abrupt drop towards the asymptotic tail.
One could describe this structure as having a thick spherical shell
with the energy density which is larger than the energy density in the core.
It is worth noting that the dominant energy condition,
which requires that the energy density is non-negative,
and that it is greater than
or equal to the absolute values of any of the individual pressures,
is satisfied in all solutions we have examined.
One also observes that the mass function comes close to its asymptotic value
(total mass $M$) well before the spatial infinity is reached,
as well as the maximum of the compactness function,
which gives a measure of the effective size
of the self-gravitating object formed by the scalar field.
The right panel of figure\ \ref{fig:threefour} shows the close-up view
of the asymptotic tail of the field profile function,
the energy density and the pressures.
Qualitatively, the situation is no different
from what one finds in the standard theory
(see~\cite{liddlemadsen}, p.\ 114);
the energy density and the radial pressure approach zero from above,
while the transverse pressure crosses zero,
reaches its minimum, and approaches zero from below.
It is important to emphasize that, within the present model,
the components of the energy--momentum tensor
approach the vacuum state only asymptotically,
implying that there is no possibility of joining the interior spacetime
with the exterior vacuum spacetime at some finite $r$.%
\footnote{By using the Israel's junction surface formalism,
it follows that in order to join the two spacetimes
at the hypersurface $r=R$ (without introducing the $\delta$-shell
energy--momentum distribution on the hypersurface itself)
the radial pressure must be continuous across the hypersurface,
while the energy density and/or the transverse pressure can be discontinuous.
As the vacuum solution has $p=0$,
this would require $p\to0$ as $r$ approaches $R$ from the inside.
This feature is not found in our solutions.}
As a consequence, there is no strict definition of the stellar radius.
%

%%%%%%%%%%%%%%%%%%%%%%%%%%%%%%%%%%%%%%%%%%%%%%%%%%%%%%%%%%%%%%%%%%%%%%%%%%%%%%

\section{Conclusions \label{sec:concl}}

While in the TEGR 
any choice of the tetrad leads to the same equations of motion,
this is not necessarily so if modifications to the theory are introduced.
Using the language \cite{tamanini}
which deals with the $f(T)$ extension of the theory,
one can speak of `good' and of `bad' tetrads,
depending on the structure of the resulting equations of motion.
In this paper we provide a clear example of these concepts.
We have investigated one of the simplest matter models, the scalar field,
in one of the simplest geometrical settings, that of spherical symmetry.
We included the coupling of the scalar field to the torsion scalar
in a way that resembles the widely studied
non-minimal coupling of the scalar field to the curvature scalar.
The algebraically simplest choice of the tetrad
resulted in the equations of motion
that evidently could not have a solution
(except if field-to-torsion coupling was removed).
Trying out different tetrads, one particular tetrad was found
for which we obtained the self-consistent set of equations.
These equations could be solved numerically
and a spectrum of configurations could be examined in detail.
It is however possible that a different `good' tetrad,
leading to different set of self-consistent equations of motion,
and consequently to boson stars with different properties, exists.
In this sense, our results can be considered as tetrad-specific.
Having found at least one `good' tetrad
can certainly be seen as a success,
but we must remain aware that it was found by trial and error
(or `by accident'), and not through the application of a method
that could be useful in similar circumstances as well.
In principle, the method could consist of
writing down the general ansatz for a tetrad,
which would involve six spacetime-dependent functions
representing the parameters
of the local Lorentz transformation of the tetrad,
and using the resulting equations of motion to
single out the `good' tetrad.
Such a general procedure is still out of our hands.

The main result of this work is a new class of boson stars
with interesting physical properties.
We have constructed self gravitating objects
formed by the complex scalar field nonminimally coupled to torsion.
All configurations we have considered involve anisotropic principal pressures
which obey the dominant energy condition.
In the configurations with sufficiently large
field-to-torsion coupling constant $\xi$
we have found the increasing energy density
as one moves away from the centre of the star which,
after reaching its maximum at a finite radial distance from the centre,
suddenly drops to the usual asymptotic tail.
This feature could be described as a thick shell
around the core of the star which has lesser energy density than the shell.
The radius at which the energy density has the maximum
can be taken as the measure of the size of these objects since,
as we are dealing with extended, and not with compact objects,
there is no strict definition of their radius.
The shell was obtained in configurations which are,
on the basis of the analogy with the stability properties
of the boson stars with minimally coupled field,
expected to be dynamically stable.
This is not the first indication of the possible dynamical stability
of bodies with outwardly increasing energy density.
For example, thick shells were also found in bodies
constructed with some quasi-local equations of state
of the anisotropic fluid, which were shown to be stable \cite{horvat-nleos}.
However, adequate analysis is required to prove or disprove
the stability of the boson stars constructed in this paper,
and as the equations of motion are not much more
complicated than those of the minimally coupled field,
the perturbative approach seems to be the most direct route.
Another extension of the present work
could be the inclusion of field self-interaction
in the form of the $\phi^4$ term, or more general potential terms,
since these may have significant effect on the structure
of the self-gravitating bodies (for the effects of the potential terms
within the standard curvature theory see e.g.~\cite{coshawa}
which regards the masses of the boson stars,
or \cite{compactbstar} where a V-shaped potential
makes it possible to construct compact boson stars).
It would also be interesting to investigate
whether even more exotic spherically symmetric structures
than the ones obtained here,
e.g.~gravastars \cite{visseranisograva,adebenedgravastar},
could be supported by the scalar field nonminimally coupled to torsion.

%%%%%%%%%%%%%%%%%%%%%%%%%%%%%%%%%%%%%%%%%%%%%%%%%%%%%%%%%%%%%%%%%%%%%%%%%%%%%%

\vskip 1em

%%%%%%%%%%%%%%%%%%%%%%%%%%%%%%%%%%%%%%%%%%%%%%%%%%%%%%%%%%%%%%%%%%%%%%%%%%%%%%

\noindent {\bf Acknowledgments:}
The authors would like to thank Andrew DeBenedictis
for suggesting teleparallel gravity to us.
This work is supported by the University of Zagreb
grant nr.~I-301-00613-VIF2013-02.
Partial support comes from ``NewCompStar'', COST Action MP1304.

%%%%%%%%%%%%%%%%%%%%%%%%%%%%%%%%%%%%%%%%%%%%%%%%%%%%%%%%%%%%%%%%%%%%%%%%%%%%%%

%\bibliography{sail}
\providecommand{\href}[2]{#2}\begingroup\raggedright\endgroup

%%%%%%%%%%%%%%%%%%%%%%%%%%%%%%%%%%%%%%%%%%%%%%%%%%%%%%%%%%%%%%%%%%%%%%%%%%%%%%

\end{document}